\documentclass[review,12pt]{elsarticle}
\usepackage{float,appendix,multicol}
\usepackage[utf8]{inputenc}    % utf8 support       %!!!!!!!!!!!!!!!!!!!!
\usepackage[T1]{fontenc} 
\usepackage{amsmath,amssymb, mathtools,mathrsfs,stmaryrd,titletoc}

\usepackage{natbib}
\biboptions{sort&compress} %To make [1,2,3] be [1-3]
\usepackage[retainorgcmds]{IEEEtrantools}
\usepackage[usenames]{color}
\usepackage{tabularx}
\usepackage{booktabs}
\usepackage{psfrag}
\usepackage[font=small,labelfont=md]{caption,subfig}
\usepackage{multirow}
\usepackage[T1]{fontenc} % typing french                        
\usepackage[bookmarks=true,colorlinks=true,linkcolor=blue,citecolor=red]{hyperref}
\usepackage{float}         % make new float environment such as boxes (captioned)
\usepackage{listings}      % insert source code   
\usepackage{algorithm}
\usepackage{algorithmicx}
\usepackage{algpseudocode}
\usepackage{verbatim}
\usepackage{numprint}

\usepackage{silence}
\usepackage{siunitx}
\usepackage[norefs,nocites,ignoreunlbld]{refcheck} % warning for unreferred figs/tables/equas
\usepackage[activate={true,nocompatibility},final,tracking=true,kerning=true,spacing=true,factor=1100,stretch=10,shrink=10]{microtype}

\usepackage[capitalise]{cleveref} %Basically, cleveref must be loaded last.
\crefname{figure}{Fig.}{Figs.}
\crefname{equation}{Eq.}{Eqs.}
\crefname{section}{Section}{Sections}
\usepackage[textsize=tiny]{todonotes}
\usepackage{nicefrac} % type inline fractions: \nicefrac{1}{2}
\usepackage{setspace}
\usepackage{lineno}  % write numbers for lines
\usepackage{totcount} % to count the total number of references and other things
\usepackage[figure,table]{totalcount}
\usepackage{blkarray, bigstrut} % write complicated matrices with borders, see http://mirror.lagoon.nc/pub/ctan/macros/latex/contrib/blkarray/blkarray.pdf
\usepackage{setspace}
\usepackage{tikz}
\usetikzlibrary{arrows,decorations.pathmorphing,decorations.pathreplacing,backgrounds,positioning,fit,matrix,math,shapes.misc}
\tikzset{cross/.style={cross out, draw=black, minimum size=2* (#1-\pgflinewidth), inner sep=0pt, outer sep=0pt}, cross/.default={1pt}}
\usepackage{wasysym}
\usepackage{gensymb} % for degree symbol
\usepackage[section]{placeins} % to place all figures before a new section

\usepackage{xspace}

\newcommand{\trace}{\mathrm{tr}}
\newcommand{\bfsigma}{\boldsymbol{\sigma}}
\newcommand{\bfepsilon}{\boldsymbol{\varepsilon}}

\newcommand{\dd}{\mathrm{d}}
                          
\newcommand*{\pd}[3][]{\dfrac{\partial^{#1} #2}{\partial #3^{#1} }}  
\usepackage{longtable} % for table over multiple pages
\usepackage{lipsum} % just for dummy text
\usepackage[percent]{overpic}
\usepackage[T1]{fontenc}
\usepackage{lmodern}
\usepackage[margin=2.2cm]{geometry}% TO PLAY WITH THE MARGINS
\usepackage{upgreek}
\biboptions{numbers,sort&compress}
\journal{Engineering Failure Analysis}

\begin{document}
\begin{frontmatter}
\title{On the suitability of single-edge notch tension (SENT) testing for assessing hydrogen-assisted cracking susceptibility}
% or assessing sulfide stress cracking susceptibility or assessing hydrogen-assisted cracking susceptibility in sour environments 

\author[ICL]{Livia Cupertino-Malheiros} %\ead{l.cupertino-malheiros@imperial.ac.uk}
\author[Oxf]{Tushar Kanti Mandal} %\ead{t.mandal@imperial.ac.uk}
\author[Vall]{Florian Th\'{e}bault}
\author[Oxf]{Emilio Mart\'{\i}nez-Pa\~neda\corref{cor1}} 
\ead{emilio.martinez-paneda@eng.ox.ac.uk}

\address[ICL]{Department of Civil and Environmental Engineering, Imperial College London, London SW7 2AZ, UK}
\address[Oxf]{Department of Engineering Science, University of Oxford, Oxford OX1 3PJ, UK}
\address[Vall]{Vallourec Research Center France, 60 Route de Leval, 59620, Aulnoye-Aymeries, France}

\cortext[cor1]{Corresponding author.}

\begin{abstract}
Combined experiments and computational modelling are used to increase understanding of the suitability of the Single-Edge Notch Tension (SENT) test for assessing hydrogen embrittlement susceptibility. The SENT tests were designed to provide the mode I threshold stress intensity factor ($K_{\text{th}}$) for hydrogen-assisted cracking of a C110 steel in two corrosive environments. These were accompanied by hydrogen permeation experiments to relate the environments to the absorbed hydrogen concentrations. A coupled phase-field-based deformation-diffusion-fracture model is then employed to simulate the SENT tests, predicting $K_{\text{th}}$ in good agreement with the experimental results and providing insights into the hydrogen absorption-diffusion-cracking interactions. The suitability of SENT testing and its optimal characteristics (e.g., test duration) are discussed in terms of the various simultaneous active time-dependent phenomena, triaxiality dependencies, and regimes of hydrogen embrittlement susceptibility.\\  

\end{abstract}

\end{frontmatter}

\date{\today}
%------------------------------------------------------------------------------------------------------------------

%=============================================
\section{Introduction}
\label{sec:intro}

An accurate determination of the mode I threshold stress intensity factor ($K_{\text{th}}$) is essential for the safe and cost-effective design of metallic structures exposed to environments prone to hydrogen absorption. Its determination is, however, not straightforward, and better experimental and numerical methodologies for obtaining $K_{\text{th}}$ values representative of service conditions are increasingly being investigated \cite{CupertinoMalheiros2022,Xing2022,Nguyen2024}. The main challenges are intrinsically related to the mechanism of hydrogen-assisted cracking \cite{Gangloff2003,Djukic2019,Campos2024}, as the high sensitivity of the fracture toughness to the hydrogen content makes global quantities such as $K_{\text{th}}$ highly dependent on the local hydrogen concentration around the crack tip \cite{AM2016,Corsinovi2023}. This crack tip hydrogen concentration is in turn linked to the hydrogen absorbed at the metal surface and to stress-driven hydrogen diffusion within the crystal lattice. While failure often takes place under small scale yielding conditions and a remote $K$-field exists, the non-uniform distribution of hydrogen near cracks and other stress concentrators results in a non-homogeneous spatial \textit{and} temporal distribution of the fracture toughness of the metal. As a result, a global fracture description of hydrogen-assisted failures must be accompanied by an appropriate understanding of the local conditions, which involves resolving hydrogen ingress for a given environment \cite{cupertino2024hydrogen}, the interplay between local mechanical and hydrogen content states ahead of the crack tip \cite{AM2020}, and the time scales and dependencies of all the phenomena involved \cite{Hageman2023b}. Hence, the development of new testing methodologies and the assessment of their suitability for characterising hydrogen embrittlement susceptibility require resolving the coupling between global ($K_{\text{th}}$) and local (crack tip hydrogen content) quantities.\\ 

In recent years, there has been growing interest in the use of Single Edge Notch Tension (SENT) specimens for fracture toughness testing \cite{Furmanski2022,Turkalj2019,Alvarez2020}. This testing configuration results in lower levels of crack tip constraint (triaxiality) relative to other specimen geometries commonly used in fracture testing, such as Compact Tension (CT) and Single Edge Notch Bend (SENB) samples. Since the apparent fracture toughness increases with decreasing stress triaxiality, the SENT test is often deemed to be relatively less conservative, yet it provides a better representation of the crack tip constraint levels relevant to the growth of axial cracks from the inner to the outer diameter of the pipeline walls \cite{Dadfarnia2011b,Panico2017,Turkalj2019}. Accordingly, due to the mechanical and environmental similitude between SENT samples and cracked pipelines \cite{Dadfarnia2011b}, the use of SENT data can reduce conservatism, enabling more realistic predictions of pipe burst pressures compared to full-scale tests in acidic corrosive environments \cite{Turkalj2019}. However, despite ongoing efforts to understand and improve the SENT method for hydrogen-assisted cracking, the interactions between test conditions (specimen geometry, test duration), environmental severity, and cracking are still not adequately understood \cite{Li2018d}. This lack of fundamental understanding can hinder the applicability of the test to obtain reliable $K_{\text{th}}$ values and plague the test protocol with uncertainties.\\

In this work, constant-load SENT tests were designed to provide the $K_{\text{th}}$ for hydrogen-assisted cracking of a C110 low-alloy steel exposed to two corrosive environments. These tests were
accompanied by hydrogen permeation experiments to relate environmental conditions to hydrogen absorption. A phase field-based,
coupled deformation-diffusion-fracture finite element model is then employed to model the SENT tests, predicting $K_{\text{th}}$ values in good agreement with the experimental findings. The experimental and numerical results obtained are discussed in relation to the hydrogen absorption-diffusion-cracking interplay, bringing new key insights that contribute to identifying suitable conditions for SENT testing in corrosive environments.  

%=============================================
\section{Experimental methods}
\label{sec:experiment}
%=============================================

%++++++++++++
\subsection{Material}
\label{SubSec:Material}
%++++++++++++
The material employed was a C110 low-alloy steel grade (Fe-0.3C-0.3Si-0.4Mn-0.6Cr-1.0Mo 0.1V-0.03Nb weight\%), extracted from a tube with 356 mm outer diameter and 30 mm wall thickness supplied by Vallourec Research Center France. The steel presents a fully tempered martensitic microstructure. Uniaxial tensile testing in the air at a strain rate of $10^{-4}$ s$^{-1}$ revealed an engineering yield strength and ultimate tensile strength of 820 and 883 MPa, respectively, and a fracture strain of 19\%.

%++++++++++++
\subsection{Hydrogen Permeation}
%++++++++++++

Permeation data is acquired using a Devanathan-Stachurski \cite{Devanath1962} two-compartment electrolytic cell at 24 $^\circ$C. Permeation transients are recorded for 720 hours using a Bio-Logic VMP3 potentiostat, Ag/AgCl reference electrode and platinum mesh counter electrode. The permeation membranes tested have a thickness of 2.9 ± 0.1 mm and an exposed area of 6.6 $\text{cm}^2$ (sample holder with 2.89 cm diameter opening). Following the grinding of both membrane sides using SiC papers up to 1200 grit (15 $\upmu$m), the exit surface was electroplated with a thin layer of Pd and maintained at a constant anodic potential of 0.33 $\text{V}_\text{Ag/AgCl}$ in a 0.1 mol/L NaOH solution. After the anodic current becomes stable and below 200 nA/cm\textsuperscript{2}, an acidic solution (5.0 wt\%NaCl, 1-2 wt\%CH$_3$COOH, pH $\approx$ 3) saturated with a gas containing 3, 10 or 100\% H$_2$S is poured into the charging compartment of the cell. The anodic solutions were deaerated with nitrogen one hour before and during the experiments. The charging solutions were first deaerated and then saturated with the respective gas for 45 min before being poured into the charging compartment. The bubbling of the charging gas was then maintained throughout the experiment. 

%++++++++++++
\subsection{Constant-load SENT}
\begin{figure}[!htb]
    \centering
    \includegraphics[width=1\textwidth]{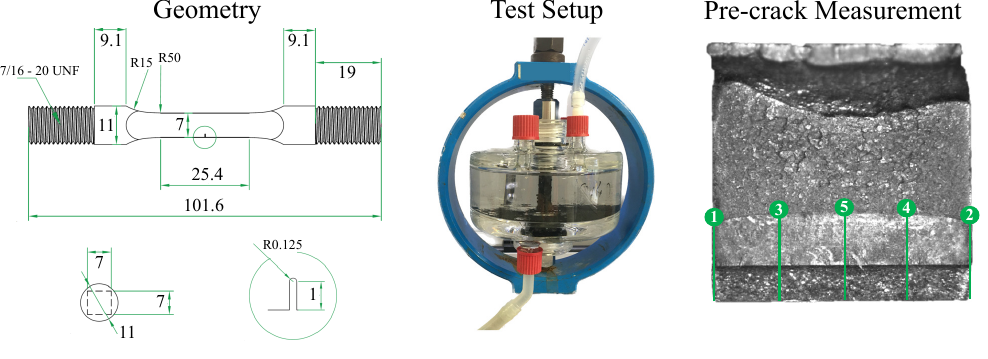}
    \caption{SENT specimen configuration: geometry (left), with dimensions in mm, picture of the specimen under constant load using a proof ring (centre), and optical microscope image of fractured SENT specimen indicating the 5 equidistant locations where the pre-crack measurements are taken (right).}
    \label{fig:SENTsetup}
    \end{figure}

%++++++++++++
The SENT specimens tested in this work have a squared cross-section, with a width ($W$) and thickness ($B$) equal to 7 mm, and a gauge length of 25.4 mm, as can be seen in Fig. \ref{fig:SENTsetup}. They were sampled at the mid-thickness of the tube and a 1 mm notch was machined closer to the inner side of the tube wall. After machining, the specimens were ground with SiC paper up to  1200 grit (15 $\upmu$m) and pre-cracked by fatigue using the 3-point bending method. The SENT tests were performed in the same acid solution as the permeation experiments, saturated with 7\% or 100\% H$_2$S. A constant load between 7 and 14 kN was applied to each tested specimen using a proof ring (shown in Fig. \ref{fig:SENTsetup}), providing a wide range of applied $K_{\text{I}}$. The tests were stopped after the failure of the sample or, in the absence of fracture, after 720 hours. The specimens that did not fail in 720 h were fractured by 3-point bending in air shortly after the test. The fracture surface of each specimen was then examined under an optical microscope to measure the value of the initial pre-crack length ($a_0$), considered as $a_0 = 0.25(0.5 \sum_1^2 a_i + \sum_3^5 a_i)$, where $i$ corresponds to the equidistant measurements indicated in Fig. \ref{fig:SENTsetup}. In total, 12 and 10 specimens were tested with 7\% and 100\% H$_2$S, respectively, with $a_0$ between 2.03 and 2.89 mm  ($a_0/W \approx 0.3 \text{ to } 0.4$).\\

%-------------
%-------------

%=============================================
\section{Numerical methodology} 
\label{sec:formulation}
%=============================================
The hydrogen-assisted failure of the SENT samples is idealised as a coupled deformation-diffusion-fracture problem, where the evolving fracture interface is simulated using the phase field method. The characteristics of phase field fracture and its connection to Griffith's energy balance and the thermodynamics of fracture are outlined in \cref{sec:variational}. Then, in \cref{sec:elastoplastic-frac}, the model is particularised to the description of hydrogen-assisted fracture in elastic-plastic solids. This includes the definition of a suitable driving force for fracture, accounting for the fraction of plastic work that has dissipated, and the endowment of the model with a phenomenological hydrogen degradation law for the considered steel. This is followed, in \cref{sec:hydrogen-diffusion}, by the description of the hydrogen transport model, together with its coupling with the mechanical fields and the cracking process. Finally, the characteristics of the numerical SENT model, the relevant material parameters, and the chemo-mechanical boundary conditions employed are provided in Section \ref{sec:FEA_sent}.

%---------------------------------------
\subsection{A phase field description of fracture}
\label{sec:variational}
%---------------------------------------
The phase field fracture model is based on Griffith's thermodynamics framework, which assumes that a crack will grow when a critical strain energy release rate is attained \cite{Griffith1920,Francfort1998}. Mathematically, this can be expressed as follows. In a solid $\Omega$, containing a discrete crack $\Gamma$, the balance between stored and fracture energy can be described through the following total potential energy functional,
\begin{align}\label{eq:frac2}
   \mathscr{E} = \int_{\varOmega/\Gamma} \psi (\bfepsilon) \, \dd\varOmega + \int_\Gamma G_c \, \dd\Gamma,
\end{align}

\noindent where the first term corresponds to the driving force for fracture, the energy stored in the solid that is available to create two new surfaces, and the second term represents the resistance of the material to fracture, the critical energy threshold required to create two new surfaces. Here, $\psi$ denotes the strain energy density, which is a function of the strain tensor $\bfepsilon$, and $G_c$ is the so-called critical energy release rate, also referred to as fracture energy or material toughness. To enable the computational minimisation of Eq. (\ref{eq:frac2}), the discrete crack can be smeared using an auxiliary phase field variable $\phi$, such that $\phi=0$ denotes intact (undamaged) material points and $\phi=1$ represents fully cracked material points, akin to a damage variable. Then, upon suitable choices of crack density function $\gamma (\phi, \ell)$ and degradation function $g(\phi)$, the functional $\mathscr{E}$ can be regularised as,
\begin{align}\label{eq:frac3}
   \mathscr{E}_\ell = \int_{\varOmega}  \big( g (\phi ) \psi_0 (\bfepsilon)  +  G_c  \gamma (\phi, \ell) \big) \, \dd\varOmega
\end{align}

\noindent enabling the computational prediction of cracking as an exchange between stored and fracture energy. In Eq. (\ref{eq:frac3}), $\psi_0$ denotes the undamaged strain energy density and $\ell$ is the phase field length scale parameter, which ensures mesh objectivity in the computational results. Importantly, the choice of $\ell$ also determines the material strength. For example, for the following choice of crack density function,
\begin{equation}
    \gamma(\phi,\ell) = \left(\dfrac{\phi^2}{2\ell} + \dfrac{\ell}{2} |\nabla\phi|^2\right)
\end{equation}

\noindent which corresponds to the so-called conventional or \texttt{AT2} phase field model \cite{Bourdin2008}, the solution to the 1D homogeneous problem renders a critical stress of 
\begin{equation}\label{eq:Strength}
    \sigma_c = \left( \frac{27 E G_c}{256 \ell} \right)^{1/2}
\end{equation}

Consequently, the phase field fracture model not only adequately predicts cracking when the energy release rate reaches a critical quantity ($G=G_c$, or $K_I=K_{Ic}$), in agreement with well-established fracture mechanics principles, but is also capable of capturing strength-driven failures and the sensitivity to the flaw size (see Refs. \cite{Tanne2018,PTRSA2021}). 

%-----------------------------------
\subsection{Elastoplastic hydrogen-assisted fracture}
\label{sec:elastoplastic-frac}
%-----------------------------------
We proceed to particularise the model to the fracture of an elastic-plastic solid exposed to a hydrogen-containing environment. In agreement with experimental observations, and following the work by Mart\'{\i}nez-Pa\~neda \textit{et al.} \cite{CMAME2018}, the interplay between hydrogen and fracture is introduced by making the material toughness dependent on the hydrogen concentration through the definition of a hydrogen degradation function, such that,
\begin{equation}
    G_c (C) = f (C) G_c (0)
\end{equation}

\noindent where $G_c (0)$ is the critical energy release rate in a hydrogen-free environment. The definition of $f(C)$ can follow mechanistic \cite{CMAME2018,Duda2018} or phenomenological \cite{Cui2022,Mandal2024} approaches. Here, the latter class is adopted, as the relationship between fracture toughness and hydrogen content is well-characterised for the material under consideration, a low-alloy steel (see Section \ref{sec:FEA_sent}). The toughness versus hydrogen content data is fitted with the following exponential function,
\begin{equation}
\label{eq:hydrogen-Gc}
G_{c}(C) 
= \left[ \frac{G_{c}^{\text{min}}}{G_{c} (0)}  + \left( 1 - \dfrac{G_{c}^{\text{min}}}{G_{c}(0)} \right) \exp{ \left( - q C \right)} \right] G_{c} (0)
\end{equation}
where $q$ is a fitting parameter, and $G_{c}^{\text{min}}$ is the saturation magnitude, the lowest value of $G_{c}$ in hydrogen-containing environments.\\

A standard, von Mises plasticity model is adopted to describe material deformation. Small strains are assumed and consequently, the strain tensor is estimated as $\bfepsilon:=\nabla^\mathrm{sym}\mathbf{u}$ and additively decomposed into elastic and plastic parts: $\bfepsilon = \bfepsilon_e + \bfepsilon_p$. Power law hardening is assumed, such that the flow stress equals,
\begin{equation} \label{eq:PowerLaw}
  \sigma_f(\varepsilon^p) = \sigma_y \left(1 + \dfrac{E\varepsilon^p}{\sigma_y} \right)^{N}   
\end{equation}

\noindent where $N$ is the strain hardening exponent ($0 \leq N \leq 1$), $\varepsilon^p$ is the equivalent plastic strain, and $\sigma_y$ denotes the yield stress. This results in the following strain energy density
\begin{equation}
\psi_p = \dfrac{\sigma_y^2}{E(N+1)}\left(1 + \dfrac{E\varepsilon_p}{\sigma_y} \right)^{(N+1)} .
\end{equation}

To define the interplay between fracture and plasticity we follow the work by Mandal \textit{et al.} \cite{Mandal2024}. Accordingly, and consistent with thermodynamics, only the plastic work that is not dissipated into heat is assumed to be available to drive crack advance. The strain energy density is then defined as,
\begin{equation}
    \psi :=  g (\phi) \psi_e \left( \bfepsilon_e \right)
+ \bar{ g }(\phi)\psi_p(\varepsilon_p),
\end{equation}

\noindent with the elastic strain energy density being given by $\psi_e=\frac{1}{2} \bfepsilon_e : \mathcal{C} : \bfepsilon_e =\frac{1}{2} \lambda \text{tr}^2 (\bfepsilon_e) + \mu \text{tr} (\bfepsilon_e^2) $ where $\mathcal{C}$ is the fourth-order elasticity tensor, and $\lambda$ and $\mu$ are the Lamé constants. The choice of two distinct degradation functions for the elastic and plastic parts of the strain energy density provides the flexibility needed to define a thermodynamically and variationally-consistent fracture driving force. Following Ref. \cite{Mandal2024}, we adopt,
\begin{equation}\label{eq:beta}
 g (\phi) = (1-\phi)^2,\quad
\bar{ g }(\phi) := \beta g (\phi) + (1-\beta),\quad
0\le\beta \le 1.
\end{equation}
with $\beta$ being a parameter that quantifies the fraction of plastic work that is stored in the material and is therefore available to create new fracture surfaces. Following the seminal work by Taylor and Quinney \cite{taylorLatentEnergyRemaining1933}, a value of $\beta = 0.1$ is adopted, as only $10\%$ of the plastic work does not dissipate as heat. In addition, a history field $\mathcal{H}$ is introduced to ensure damage irreversibility \cite{Miehe2010a}. Considering the constitutive choices above and taking the variation of $\mathscr{E}_\ell$ with respect to the phase field $\phi$ and displacement field $\mathbf{u}$ variables, the governing equations of the coupled elastic-plastic deformation-fracture problem can be obtained
\begin{align}\label{eq:eqn-disp}
\begin{split}
& \nabla \cdot \bfsigma = \boldsymbol{0}, \,\,\,\,\, \text{with} \,\,\,\,\, \bfsigma =  g (\phi)\mathbb{E} : \left(\bfepsilon - \bfepsilon_p \right) \,\,\,\,\, \text{and} \,\,\,\,\, ||\bfsigma_\mathrm{dev}|| = \bar{ g }(\phi)\sigma_f(\varepsilon^p)\;\;\mathrm{if}\;\dot{\varepsilon}^p > 0,\\
& G_c (C) \left( \frac{\phi}{\ell} - \ell \nabla^2 \phi \right) =2 (1 - \phi ) \mathcal{H} \,\,\,\,\, \text{with} \,\,\,\,\,  \mathcal{H} := \max_{t}{\left(\psi_{e}^{+}(t)\right)} + \beta\psi_p.
\end{split}
\end{align}

%-----------------------------------
\subsection{Hydrogen transport}
\label{sec:hydrogen-diffusion}
%-----------------------------------
It only remains to model the transport of hydrogen within the bulk metal. Hydrogen diffusion through the crystal lattice is described using the hydrogen concentration $C$ as the primary variable. Denoting $\mathbf{J}$ as the hydrogen flux, the mass balance equation reads,
\begin{equation}\label{eq:Fick}
    \pd{C}{t} = -\nabla\cdot \mathbf{J}, \quad \text{with} \quad
    \mathbf{J} = - D \nabla C + \dfrac{DC}{RT}  \bar{V}_H  \nabla \sigma_h,
\end{equation}
where $D$ is the (apparent) hydrogen diffusion coefficient, $R = 8.314$ J/(mol K) is the ideal gas constant, $T$ is the absolute temperature, $\sigma_h=\trace{\bfsigma}/3$ is the hydrostatic stress, and $\bar{V}_H$ is the partial molar volume of hydrogen in solid solution. The last two terms account for the role that lattice distortion plays in hydrogen transport as a result of mechanical straining.\\

It is also necessary to handle the transport of hydrogen through exposed (surface) cracks. It is expected that the hydrogen-containing aqueous electrolyte will immediately occupy the space created by crack advance. To capture this, the diffusivity of material points in the damaged region is significantly enhanced by means of an amplification coefficient $k_d \gg 1$. A damage threshold, $\phi_\mathrm{th} \approx 0.8$, is also defined to establish the degree of damage above which the electrolyte is assumed to progress through an interconnected network of microcracks. Then, the diffusion coefficient is reformulated as,
\begin{equation}\label{eq:artificial-diffusivity}
    D = D_0 \big( 1 + k_d\langle \phi - \phi_\mathrm{th} \rangle \big)
\end{equation}
where $D_0$ is the hydrogen diffusivity of the undamaged material and $ \langle \Box \rangle$ denote the Macaulay brackets. Alternative approaches, such as the use of penalty-based \emph{moving} chemical boundary conditions \cite{CS2020}, can also be adopted to ensure that the hydrogen concentration in damaged regions equals that of the environment, $C (\phi >0.8) \approx C_{\text{env}}$.

\subsection{Numerical SENT experiments} 
\label{sec:FEA_sent}

\begin{figure}[!htb]
    \centering
    \includegraphics[width=0.6\textwidth]{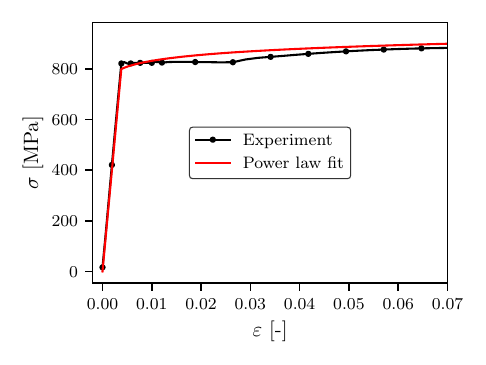}
    \caption{Uniaxial stress-strain relation: experimental data and power law fit considered in the numerical model.}
    \label{fig:uniaxial}
    \end{figure}

Finally, the model is particularised to the material and testing configuration of this work. The numerical implementation of the model is carried out in the commercial finite element package \texttt{Abaqus} using a combination of \texttt{UMAT} and \texttt{UEL} subroutines \cite{Materials2021,Mandal2024}. The coupled deformation-diffusion-fracture three-field problem is solved in a staggered manner. The material parameters of the material employed, a C110 low-alloy steel grade, are given in Table \ref{tab:database}. Common to low-alloy steels, Young's modulus and Poisson's ratio are assumed to be $E=207$ GPa and $\nu=0.3$, respectively. The yield stress and strain hardening exponent are obtained by fitting the power law hardening expression of Eq. (\ref{eq:PowerLaw}) to the output of the uniaxial stress-strain tests discussed in Section \ref{SubSec:Material}; the experimental data and the fit are shown in Fig. \ref{fig:uniaxial}. The diffusion coefficient of the material equals $D_0=1.4 \times 10^{-4}$ mm$^2$/s, as determined via electropermeation experiments (see Section \ref{Sec:permeation}). Also, common to iron-based materials, the partial molar volume of hydrogen in solid solution is taken to be $\bar{V}_H=2000\;\mathrm{mm}^3/\mathrm{mol}$.

%-------------
%-------------

%---------------------------
    \begin{table*}[htp]
    \centering
    \setlength\fboxsep{0pt}
    \smallskip%
    \renewcommand\arraystretch{1}
    \begin{tabularx}{0.7\textwidth}{lccccc}
    \toprule
    $E$ [MPa] & $\nu$ & $\sigma_y$ [MPa] & $N$ [-] & $G_{c}(0)$ [N/mm] & $D$ [mm$^2$/s]   \\
    \midrule 
    $207,000$ & 0.3 & 800 & 0.04 &  40 &  $1.4 \times 10^{-4}$  \\
    \bottomrule
    \end{tabularx}
    \caption{Material parameters characterising the mechanical and hydrogen transport behaviour of C110 low-alloy steel.}
    \label{tab:database}
    \end{table*}

\begin{figure}[!htb]
    \centering
    \includegraphics[width=0.95\textwidth]{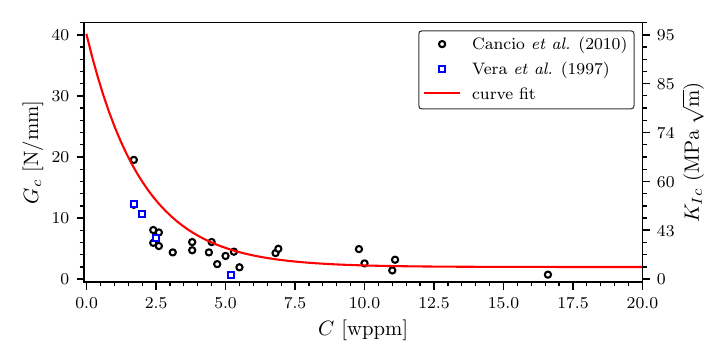}
    \caption{Sensitivity of the fracture toughness to the hydrogen concentration in low-alloy steels. $J_{Ic}$ (or $G_c$) versus $C$ data taken from Refs. \cite{Cancio2010,Vera1997}. The fitting curve corresponds to Eq. (\ref{eq:hydrogen-Gc}) with the parameters $q=0.5$ and $G_c^{\text{min}}=2$ N/mm.}
    \label{fig:database}
    \end{figure}

The fracture behaviour is described by the choices of toughness $G_c$ and phase field length scale $\ell$ (or strength $\sigma_c$, as per Eq. (\ref{eq:Strength})). The toughness, and its sensitivity to the hydrogen content, are chosen based on experimental data available for similar low-alloy steels - see Fig. \ref{fig:database}. The data, taken from Refs. \cite{Cancio2010,Vera1997}, is shown in terms of $G_c$ (or $J_c$) versus $C$, in wt ppm. The right-axis also shows the corresponding values of fracture toughness $K_{Ic}$, using the standard plane strain conversion: $K_{Ic} = \sqrt{E G_{c} / (1-\nu^2)}$. As shown in Fig. \ref{fig:database}, the experimental data can be adequately fitted with the exponential function given in Eq. (\ref{eq:hydrogen-Gc}) upon the following parameter choices: $q=0.5$ and $G_c^{\text{min}}=2$ N/mm. The phase field length scale is taken to be equal to $\ell=0.085$ mm, based on Eq. (\ref{eq:Strength}) and on the consideration of a fracture strength equal to $\sigma_c=4 \sigma_y$, as per Ref. \cite{Mandal2024}. This results in a plastic zone size $R_p=(1/3\pi)(K_{Ic}/\sigma_y)^2$ of approximately 0.15 mm, for a $K_{Ic}$ value of $\sim$30 MPa$\sqrt{\text{m}}$ (see below).\\

%-------------

\begin{figure}[H] 
    \centering
    \includegraphics[width=1\textwidth]{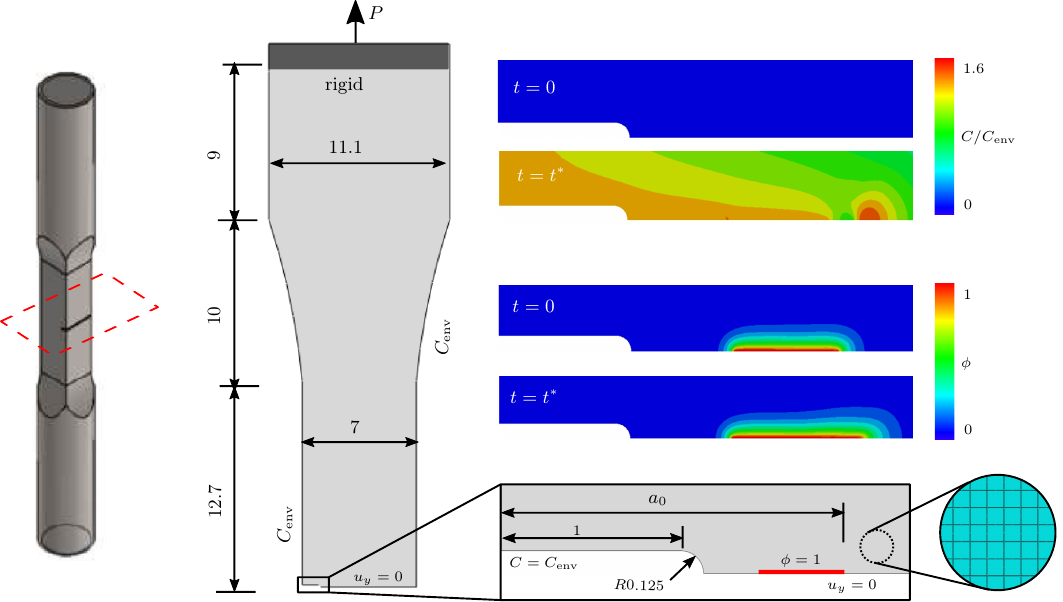}
    \caption{Geometry and numerical model of the single edge notched tensile (SENT) test. A representative distribution of the hydrogen concentration $C$ and the phase field fracture parameter $\phi$ are also shown. The initial crack, of length $a_0$, is introduced through a combination of a geometrical definition and a Dirichlet $\phi=1$ boundary condition. All geometrical dimensions are in mm.}
    \label{fig:sent-geo}
\end{figure}

The SENT geometry is modelled as a 2D plane strain problem with the dimensions given in Fig. \ref{fig:sent-geo}. Taking advantage of symmetry, only half of the specimen is simulated, with appropriate boundary conditions being used along the symmetry line ($u_y=0$). Mimicking the experiments, the loading is applied as a constant vertical load on the top edge of the sample, where only vertical translational motion is allowed. This was achieved using a rigid body constraint at the top edge; the centre point of the top edge was taken to be the reference point, where a vertical concentrated load was applied and horizontal translation and in-plane rotation were constrained. From the applied load $P$ and the crack length $a$, a mode-I stress intensity factor $K_I$ value can be obtained as,
\begin{equation}
    K_I = \dfrac{P}{B\sqrt{W}} \cdot f\left( \dfrac{a}{W}\right)
\end{equation}
where $B$ is the sample thickness and $W$ denotes its width. As per Ref. \cite{chekchaki2021procedure}, for this particular SENT testing configuration, the geometry factor $f$ is given by
\begin{equation}
  f\left( \dfrac{a}{W}\right) = 43.367\left( \dfrac{a}{W}\right)^5-99.18\left( \dfrac{a}{W}\right)^4+87.038\left( \dfrac{a}{W}\right)^3-31.762\left( \dfrac{a}{W}\right)^2+8.8764\left( \dfrac{a}{W}\right).
\end{equation}

\begin{figure}[!htb]
    \centering
    \includegraphics[width=0.8\textwidth]{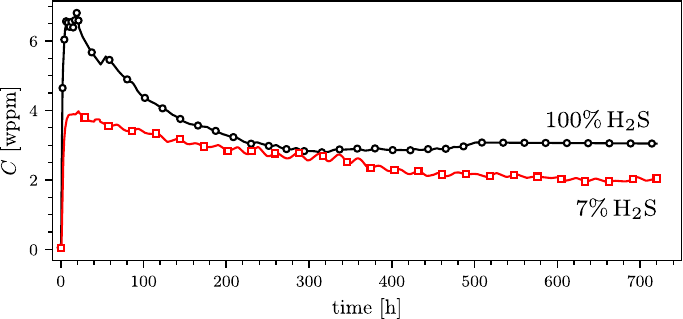}
    \caption{Time-dependent Dirichlet hydrogen concentration at the exposed boundaries of the sample, for each of the considered H$_2$S solutions. These numerical boundary conditions are based on the permeation experiments discussed in Section \ref{Sec:permeation}.}
    \label{fig:hydrogenBC}
    \end{figure}

As depicted in Fig. \ref{fig:sent-geo}, the initial crack (of length $a_0$) is defined geometrically (i.e., not enforcing the symmetry $u_y=0$ condition) and using a Dirichlet condition of $\phi=1$, as this is shown to more accurately result in an onset of crack growth for $G=G_c$ (see Refs. \cite{PTRSA2021}). The chemical boundary condition corresponds to a time-dependent hydrogen concentration $C_\mathrm{env}(t)$ that best represents the environmental conditions of the electrolyte, as determined by the permeation experiments. The outcome of those experiments, discussed in Section \ref{Sec:permeation}, results in the application of the time-dependent hydrogen surface concentration shown in Fig. \ref{fig:hydrogenBC}. \\

%-------------
%-------------

The domain is discretised using quadratic quadrilateral elements with reduced integration. A total of approximately 15,000 elements are employed, with the mesh being particularly fine along the expected crack propagation region, where the characteristic element length is at least 5 times smaller than the phase field length scale $\ell$, ensuring mesh independent results \cite{PTRSA2021}. Representative results are also given in Fig. \ref{fig:sent-geo}, in terms of the hydrogen concentration $C$ and phase field $\phi$ contours. The results exhibit the qualitative trends expected. If the applied constant load is sufficiently large, then crack growth will take place at a representative time $t^*$. Since the samples are not pre-charged, there is no hydrogen at $t=0$ but after a sufficient time the hydrogen distributes itself, with $C\approx C_{\text{env}}$ near the surface and a peak of hydrogen content in the close vicinity of the crack, coincident with the hydrostatic stress peak.

%--------------
%--------------

%================================
\section{Results} 
\label{sec:results}
%=============================================

We proceed to present the experimental results obtained for the permeation (Section \ref{Sec:permeation}) and SENT (Section \ref{Sec:fracture}) studies. These results are then subsequently extensively discussed in Section \ref{sec:discussion}. 
%---------------------------
\subsection{Hydrogen Permeation}
\label{Sec:permeation}
\begin{figure}[!htb]
    \centering
   \includegraphics[width=1\textwidth]{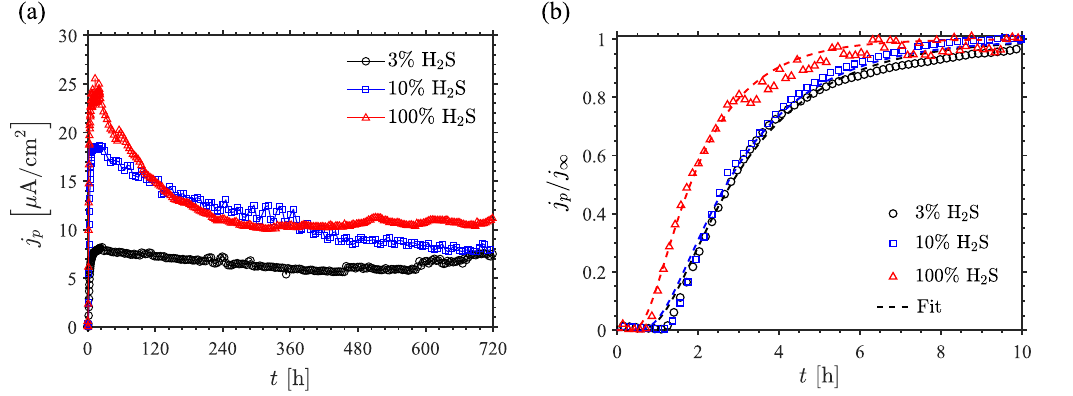}
   \caption{Hydrogen permeation transients for the three percentages of H$_2$S tested. In (a), the evolution of the anodic current density is displayed over 720 hours, while in (b) the normalized transients over the first 10 hours are shown with their respective fittings (dashed lines).}
    \label{fig:permeationAB}
    \end{figure}

%---------------------------
In this work, hydrogen uptake into the steel is the result of corrosion processes that occur on the sample surfaces that are exposed to acid environments containing H$_2$S. Electrochemical permeation is used to provide the link between the environment and the absorbed hydrogen concentration. The experimental results obtained are shown in Fig. \ref{fig:permeationAB}, in terms of the evolution in time of the anodic current density, which is proportional to the hydrogen flux at the exit side of the permeation membrane. As shown in Fig. \ref{fig:permeationAB}, three solutions were considered, 3\%, 10\% and 100\% H$_2$S, with the anodic current density being higher with increasing H$_2$S content, in agreement with expectations. The results show that there is an increase in the anodic current in the first few hours of the permeation test, reaching a steady-state current density ($j_\infty$) after approximately 10 hours. For the 10\% and 100\% H$_2$S conditions, the current density remains at this maximum $j_\infty$ value only during the first day of testing, followed by a decrease until it reaches another plateau at around 360 hours. Similar trends in permeation current density have been reported in the literature and were associated with a decrease in corrosion rates due to the formation of a protective layer of corrosion products on the steel surface \cite{Zhou2013,Zheng2014,Huang2017b}.\\

The normalised permeation transients during the first hours of the tests are shown in Fig. \ref{fig:permeationAB}(b). They can be described by the following expression,
\begin{equation} \label{eq:Fourier}
 \frac{j_p-j_0}{j_{\infty}-j_0} = 1 + 2\sum_{n=1}^{\infty} (-1)^{n} \exp {\left(-n^2\pi^2 \frac{Dt}{l^2}\right)}
\end{equation} 
considering the sub-surface hydrogen concentration at the entry side of the membrane ($C_{0}$) to be constant and given by 
\begin{equation} \label{eq:Csubsurf}
 C_0 = \frac{j_{\infty}l}{FD}
\end{equation}
where $j_p$ is the measured anodic permeation current density at a time $t$, $D$ is the apparent diffusion coefficient, $j_0$ is the initial current density, and $l$ is the membrane thickness. The experimental permeation transients were satisfactorily fitted (R$^2 \geq  0.99$) to Eq. (\ref{eq:Fourier}) using \texttt{Matlab}, resulting in $D$ values equal to 1.1, 1.2 and 1.9$\times 10^{-10}$  m$^2$/s for the 3\%, 10\% and 100\% H$_2$S solutions, respectively. The average $D$ value of 1.4$\times 10^{-10}$ m$^2$/s is then used for the $C_{0}$ calculation, Eq. (\ref{eq:Csubsurf}), and in the numerical SENT experiments, Eq. (\ref{eq:Fick}).\\ 

%-------------
%-------------

\begin{figure}[!htb]
    \centering
   \includegraphics[width=0.58\textwidth]{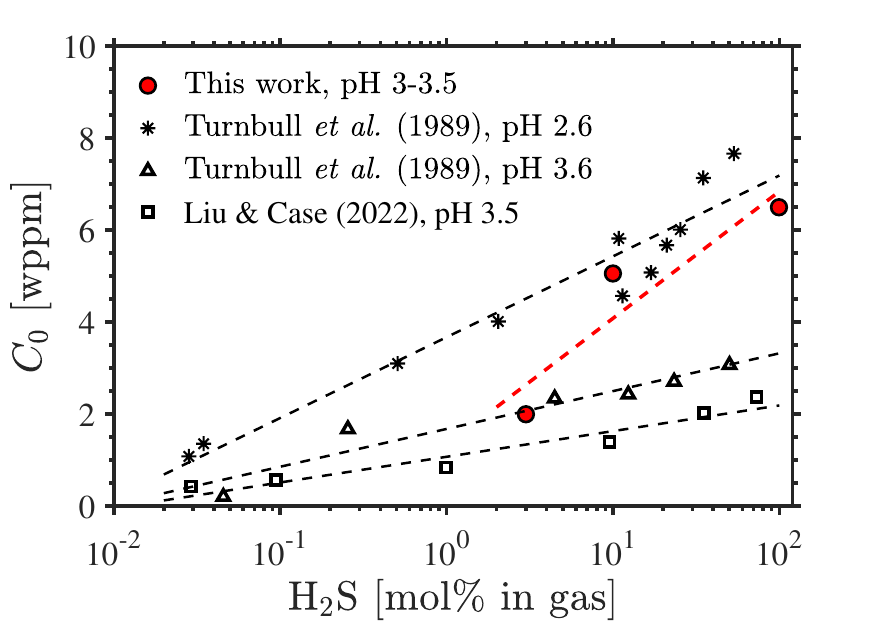}
   \caption{Sub-surface hydrogen concentration of permeation transients as a function of H$_2$S from this work, Turnbull \textit{et al.} \cite{Turnbull1989} and Liu and Case \cite{Liu2022} in NaCl solutions with pH varying from 2.6 to 3.6.}
    \label{fig:permeationC}
    \end{figure}

It has been previously reported that the sub-surface hydrogen concentrations ($C_{0}$) evolve linearly with the logarithm of H$_2$S concentration, depending on the solution pH. Fig. \ref{fig:permeationC} shows how the data from the present work compares with two previous permeation studies in similar acid NaCl solutions at room temperature \cite{Turnbull1989, Liu2022}. In the work by Liu and Case \cite{Liu2022}, the permeation tests were also performed for a C110 steel, with H$_2$S concentrations given in mol\%, while the data acquired by Turnbull \textit{et al.} \cite{Turnbull1989} used an AISI 410 steel and reported the concentrations in ppm; the equivalent mol\% in gas H$_2$S concentrations were obtained by considering an H$_2$S concentration of 2300 mg/L for 100 mol\%, as per the ANSI/NACE TM0177-2016 standard.\\ 

%-------------
%-------------

Fig. \ref{fig:permeationC} shows that our data is within the expected range of $C_{0}$ for the pH range considered, and a linear interpolation seems to be a reasonable approximation to estimate the environment hydrogen concentration ($C_\mathrm{env}$) in the H$_2$S concentration range investigated, as required for the numerical SENT tests. The time dependence related to the evolution of corrosion product films was implemented in the model by considering the decay in $j_p$ after the first day of permeation testing (Fig. \ref{fig:permeationAB}) in the calculation of $C_0$ through Eq. (\ref{eq:Csubsurf}). This approximation considers that the evolution of corrosion films on the metal surface can be represented as a series of new steady-state events and neglects the time for the anodic current density to reflect these new steady-states (a few hours). For the 7\% H$_2$S numerical results, the linear interpolation between the log of the H$_2$S concentrations of 3\% and 10\% is taken as $C_\mathrm{env} (t)$.

%---------------------------
\subsection{Experimental and numerical SENT tests}
\label{Sec:fracture}
%---------------------------

\begin{figure}[!htb]
    \centering
    \includegraphics[width=1\textwidth]{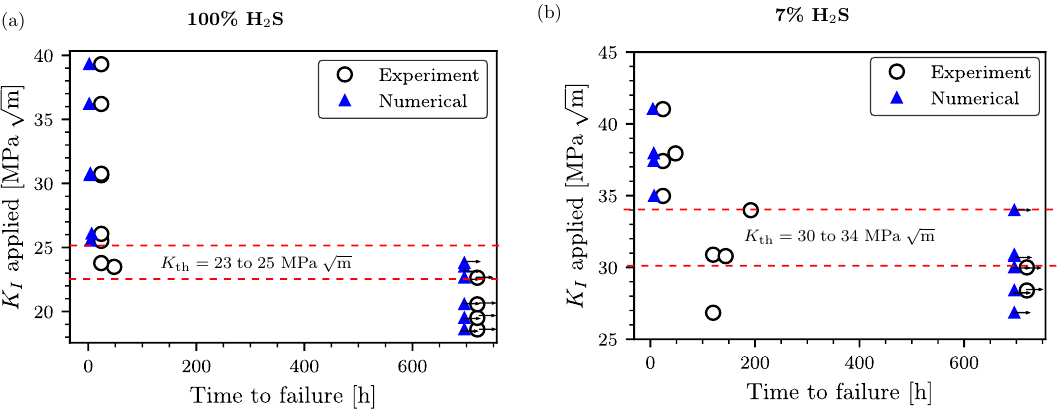}
    \caption{Experimentally and numerically determined time to failures as a function of the applied stress intensity factor $K_I$. Results are shown for two environments, based on their H$_2$S content: (a) 100\% H$_2$S, and (b) 7\% H$_2$S. For both experimental and numerical data, arrows are used to indicate that the sample did not fail within the time considered (720 h).}
    \label{fig:SIF-fail}
    \end{figure}

The experimental (circles) and numerical (triangles) time-to-failure results for all the tested SENT specimens as a function of the applied $K_I$ are shown in Fig. \ref{fig:SIF-fail}. The numerical and experimental data points represented at 720 hours correspond to the specimens that did not fail during the test, as denoted by arrows. Two scenarios are considered in terms of the environment, an aggressive 100\% H$_2$S solution and a milder 7\% H$_2$S solution. Consider first the former, Fig. \ref{fig:SIF-fail}a. Overall, a very good agreement is observed between experimental measurements and numerical predictions. The range of applied $K_I$ considered is well below the hydrogen-free fracture toughness ($K_{Ic}=95$ MPa$\sqrt{\text{m}}$) and consequently insufficient to trigger fracture at the beginning of the test, when the hydrogen has not had time to ingress and redistribute within the sample. However, as observed in the numerical model, hydrogen promptly accumulates near the crack tip, where the hydrostatic stress is highest, bringing down the material fracture resistance and triggering failures at $K_I$ levels well below the fracture toughness in air. In all cases, fracture is observed to occur before 48 h and this can be rationalised by hydrogen diffusion time scales, as elaborated in the Discussion section below (Section \ref{sec:discussion}). A distinct critical value of $K_I$ can be identified, below which failure is not observed within 720 h. Experimentally, this value is found to be around 23 MPa$\sqrt{\text{m}}$, while the computational experiments render a $K_{\text{th}}$ value of 25 MPa$\sqrt{\text{m}}$.\\ 

%-------------
%-------------

The results obtained for the 7\% H$_2$S do not exhibit the consistency of the 100\% H$_2$S condition, see Fig. \ref{fig:SIF-fail}b. First, the experimental results reveal significant scatter. A sample loaded at $K_I=30$ MPa$\sqrt{\text{m}}$ did not fail after 720 h yet failure after approximately 100 h is observed in a sample subjected to $K_I=27$ MPa$\sqrt{\text{m}}$, hindering $K_{\text{th}}$ determination. The simulations provide a precise value of $K_{\text{th}}$ but differences with the experimental data can be noted; while the finite element predictions suggest $K_{\text{th}} \approx 34$ MPa$\sqrt{\text{m}}$, the experimental data suggests a magnitude of $K_{\text{th}} \approx 30$ MPa$\sqrt{\text{m}}$, with a sample failing at a load level below that. Moreover, all the samples fracturing in the numerical experiments do so relatively early (within the first 24 h), while some of the laboratory experiments show failures after almost 200 h. These differences are extensively discussed below.

%=============================================
\section{Discussion}
\label{sec:discussion}
%=============================================

The results obtained provide interesting insight into the phenomena underpinning the hydrogen-assisted failure of SENT samples in corrosive environments. Several aspects shall be particularly emphasised.\\

\begin{figure}[!htb]
    \centering
    \includegraphics[width=1\textwidth]{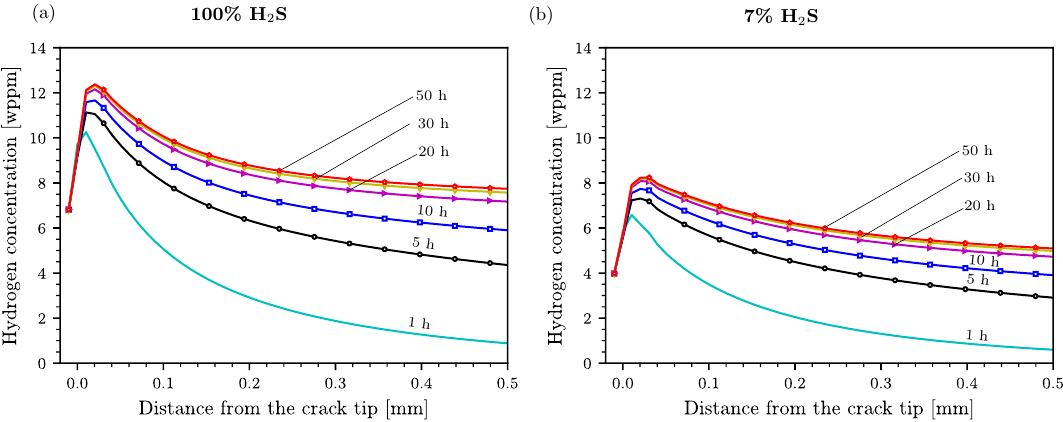}
    \caption{Distribution of hydrogen concentration ahead of the crack tip: evolution in time under the assumption of negligible protection from the corrosion product layer ($C(t) = C_\mathrm{max}$). These representative results are shown for an applied $K_I$ that corresponds to the critical threshold: (a) $K_{\text{th}} =25$ MPa$\sqrt{\text{m}}$ (100\% H$_2$S), and (b) $K_{\text{th}} = 35$ MPa$\sqrt{\text{m}}$ (7\% H$_2$S).}
    \label{fig:cracktip-Cmax-time}
    \end{figure}

\noindent \textbf{Time scales}. A key aim is to establish an optimal test duration: how soon can the test be stopped with the certainty that sample failure would not have happened later? Several competing time-dependent phenomena take place. First, common to both the aggressive (100\% H$_2$S) and mild (7\% H$_2$S) environments is the progressive formation of a corrosion product layer that hinders hydrogen ingress - as shown in Fig. \ref{fig:hydrogenBC}, the maximum sub-surface hydrogen concentration levels are attained after approximately 24 hours, and subsequently the hydrogen content drops progressively. Secondly, there is a time scale associated with the diffusion of hydrogen (i.e., how long does it take for a peak hydrogen content to be attained). To determine this, we run subsequent coupled deformation-diffusion finite element calculations for the case of a stationary crack (i.e., no phase field). Several scenarios have been considered. First, as shown in Fig. \ref{fig:cracktip-Cmax-time}, we assume that the sample is exposed to a constant hydrogen concentration, the maximum attained in the permeation test. This corresponds to the assumption of a non-protective surface film, as it might be expected in a highly strained crack surface. The results show that, for both H$_2$S environments, 10 h suffice to achieve a maximum hydrogen content that is 90\% of the steady state magnitude. It can also be seen that the local hydrogen content is roughly 2 times larger than that of the subsurface, due to lattice dilation. However, the precise figure is dependent on the material model adopted (here, conventional von Mises plasticity), with more accurate strain gradient plasticity-based models predicting a higher increase in hydrogen content \cite{IJHE2016}.\\
% Review, cite \cite{IJHE2016,JMPS2019}

%-------------
%-------------

\begin{figure}[!htb]
    \centering
    \includegraphics[width=1\textwidth]{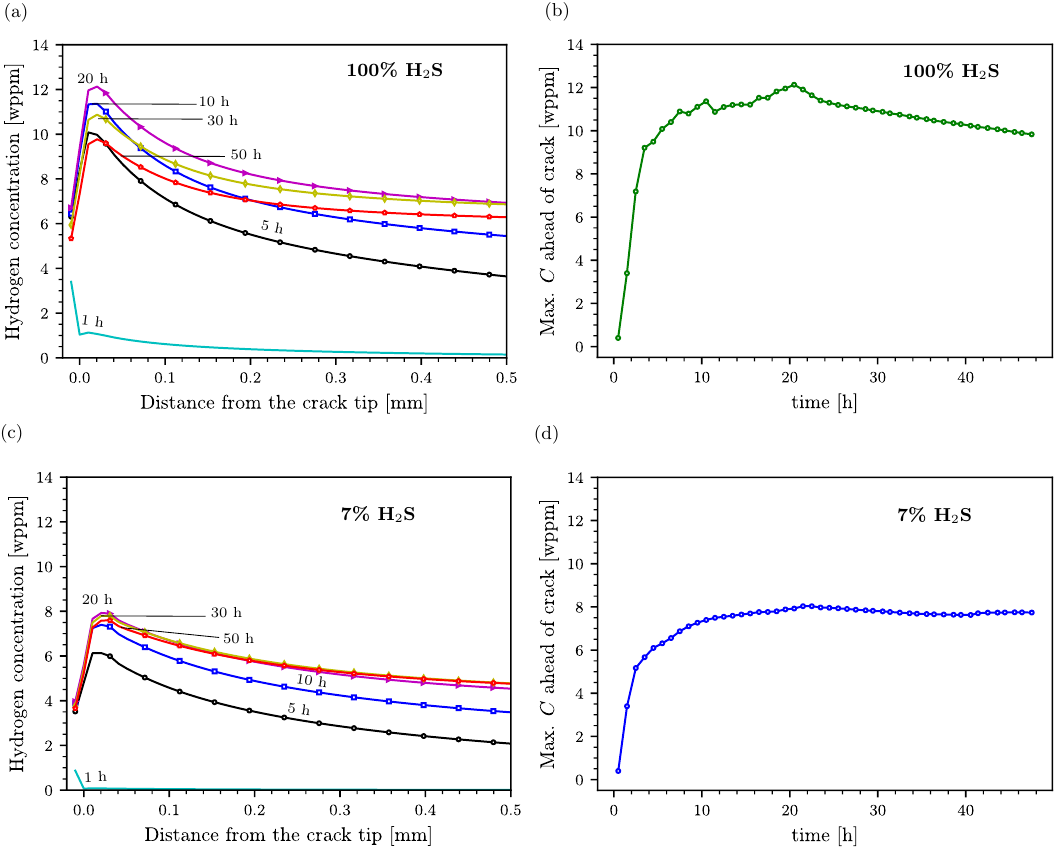}
    \caption{Hydrogen concentration ahead of the crack tip. $C$ distribution as a function of the distance to the crack tip for selected times, (a) and (c), and evolution in time of the magnitude of the peak hydrogen concentration, the maximum value of $C$ ahead of the crack tip, (b) and (d). The results have been obtained considering the role of the corrosion products in hindering hydrogen uptake, following the permeation experiments (Fig. \ref{fig:hydrogenBC}). These representative results are shown for an applied $K_I$ that corresponds to the critical threshold: $K_{\text{th}} =25$ MPa$\sqrt{\text{m}}$ (100\% H$_2$S) for (a)-(b), and $K_{\text{th}} = 35$ MPa$\sqrt{\text{m}}$ (7\% H$_2$S) for (c)-(d).}
    \label{fig:cracktip-time}
    \end{figure}

Hydrogen distributions ahead of stationary cracks are also computed considering the effect of the corrosion product layer, as shown in Fig. \ref{fig:hydrogenBC}. As in Fig. \ref{fig:cracktip-Cmax-time}, these results are shown for a given choice of applied $K_I$ ($K_I=K_{\text{th}}$), yet are quantitatively representative as the applied load only has a small effect on the peak hydrogen content (e.g., considering $K_I$ values 5 MPa higher or lower than $K_{\text{th}}$ brings less than a 10\% change in the maximum $C$ attained for the 7\% H$_2$S case). For each of the H$_2$S environments considered, results are shown in terms of both hydrogen distribution ahead of the crack tip for selected periods (Figs. \ref{fig:cracktip-time}a and \ref{fig:cracktip-time}c), and in terms of the maximum hydrogen content attained ahead of the crack tip versus time (Figs. \ref{fig:cracktip-time}b and \ref{fig:cracktip-time}d).\\

%-------------
%-------------

Consistent with the permeation data, see Fig. \ref{fig:hydrogenBC}, the peak in hydrogen content is seen to drop after 22 h, which is longer than what it takes to achieve a hydrogen distribution close to the steady state one (as per Fig. \ref{fig:cracktip-Cmax-time}). Consequently, it can be argued that the time scale that governs the attainment of the maximum crack tip hydrogen concentration is the diffusion time scale, and not the corrosion product layer formation. This significantly simplifies the analysis as correlating corrosion product formation characteristics in permeation tests and in the occluded and heavily strained conditions of cracks is not straightforward \cite{Hageman2023b}. Our numerical analysis therefore suggests that the critical hydrogen content can be achieved after approximately 10 h, potentially establishing an optimal duration of the test. However, while this is in good agreement with the computational results, some laboratory experiments show much longer failure times, particularly for the milder H$_2$S environment. As discussed below, this is likely related to stable crack growth phenomena. \\

\begin{figure}[H] 
    \centering
    \includegraphics[width=1\textwidth]{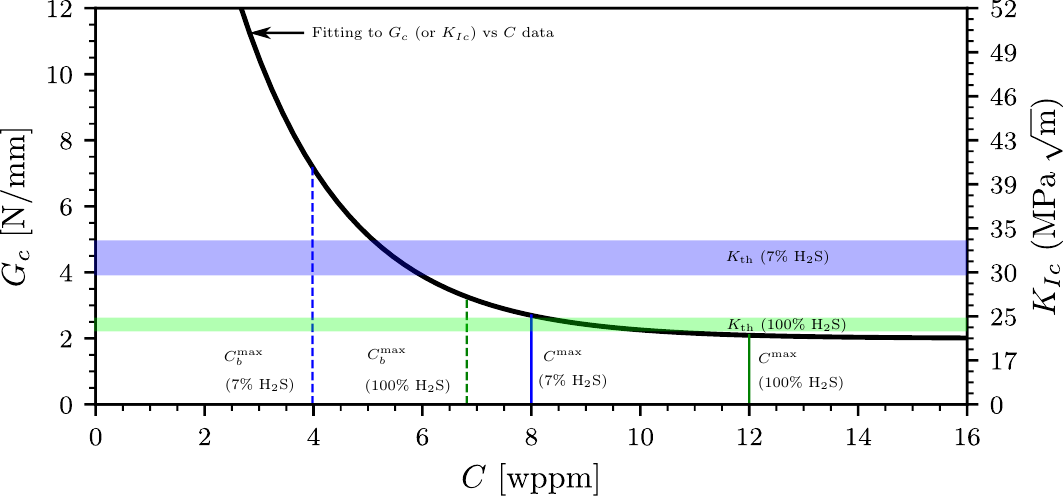}
    \caption{Rationalising the differences in $K_{\text{th}}$ determination observed for the two different environments considered: 7\% and 100\% H$_2$S. The black line denotes the fit to the toughness versus environmental (global) hydrogen content data, as obtained through DCB testing in Refs. \cite{Cancio2010,Vera1997} (see Fig. \ref{fig:database}). Vertical dashed lines are used to represent the maximum environmental hydrogen concentration at the boundary $C_b^\mathrm{max}$ for each of the solutions, as shown in \cref{fig:hydrogenBC}. Solid dashed lines are used to represent the peak hydrogen levels attained locally ahead of the crack tip, which are about 2 times higher than $C_b^\mathrm{max}$ due to stress-induced diffusion. The range of threshold $K_{\text{th}}$ values identified from our SENT experiments is also shown.}
    \label{fig:Kth-summary}
\end{figure}

\noindent \textbf{Influence of the environment}. The SENT results obtained (Fig. \ref{fig:SIF-fail}) reveal notable differences between the two environments considered: an aggressive 100\% H$_2$S solution and a milder 7\% H$_2$S one. The latter showed a higher dispersion in the experimental data, hindering the definition of $K_{\text{th}}$. These differences can be rationalised considering the fit to the toughness versus hydrogen content data, as sketched in Fig. \ref{fig:Kth-summary}. Consider first the 100\% H$_2$S solution, for which the highest value of the subsurface hydrogen content was found to be $C_b^{\text{max}} \approx 7$ wt ppm, increasing up to 12 wt ppm ahead of the crack tip. These hydrogen contents are sufficient to reduce the toughness to values very close to their saturation one (the minimum $K_{\text{th}}$ that can be obtained in a hydrogen-containing environment). For high hydrogen concentrations (8 wt ppm or larger), the shape of the toughness vs $C$ plot is reaching a plateau and consequently, results are rather insensitive to small variations in concentration. This is the opposite of what is observed in the 7\% H$_2$S case, where the hydrogen content is lower and consequently one lies in the higher slope region of the toughness versus hydrogen content curve. As a result, small changes in hydrogen content are more likely to cause changes in the $K_{\text{th}}$ measured. This would rationalise the higher degree of scatter observed for milder environments. Small test-to-test changes in surface conditions, defect geometry, or local electrolyte characteristics can result in non-negligible differences in the failure load for such low hydrogen contents. \\

%--------------

\begin{figure}[!htb]
    \centering
    \includegraphics[width=1\textwidth]{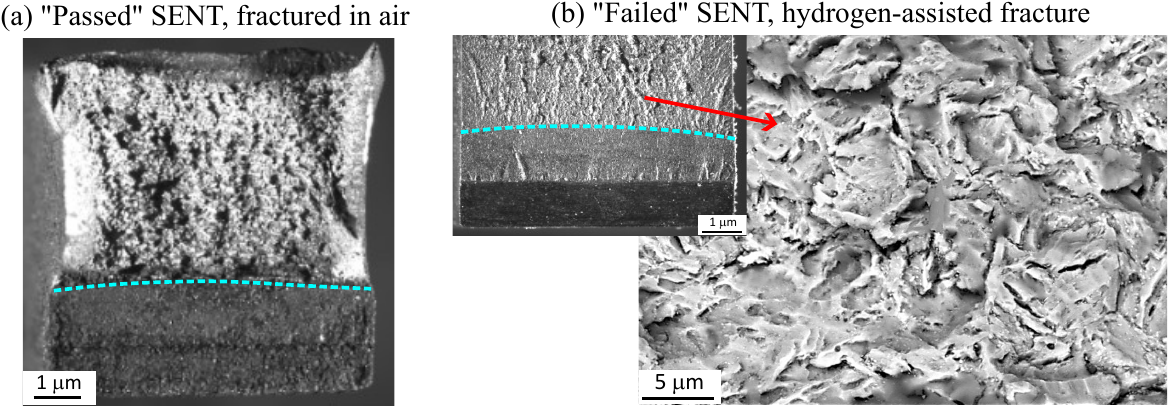}
    \caption{Representative images for the fracture surfaces of specimens that (a) passed the 720-hour SENT test and were fractured in air and (b) failed the SENT test, exhibiting hydrogen-assisted quasi-cleavage fracture morphology. The dashed blue lines delimit the end of the fatigue pre-crack.}
    \label{fig:fracturesurf}
    \end{figure}

\noindent \textbf{Opportunities for optimised and \textit{virtual} testing}. The combination of experiments and modelling has shown to be very useful in providing insight into the suitability and limitations of the SENT test for providing suitable $K_{\text{th}}$ measurements. This information can be used to optimise testing protocols and replace (or complement) laboratory testing with computational experiments. However, some challenges remain to be addressed. While the agreement between experiments and modelling is excellent for the more aggressive solution (100\% H$_2$S), differences are more pronounced for milder solutions where small changes in local hydrogen content can lead to significant variations in failure load. Moreover, as shown in Fig. \ref{fig:SIF-fail}b, some samples are observed to fail in the lab at much longer times than in the equivalent computational model, and at much longer times than those needed to achieve steady state hydrogen distributions, compromising a simple determination of optimal test duration. The data scatter and the late failures could be related to the already discussed higher toughness sensitivity to the hydrogen content in the low $C$ regime. Another potential explanation relates to subcritical crack growth. While failures can be unstable in the most aggressive environments, for mild environments a propagating crack encounters areas of increasing toughness, which could lead to crack arrest and blunting. The crack would only resume propagation once sufficient hydrogen has diffused to the new crack tip process zone, introducing another time dependency into the problem. This is observed in some of the numerical simulations, yet failure still takes place earlier than in the experiments as the time needed for the critical hydrogen content to be attained is short. This is inherently related to the homogenised description provided by the model, as the hydrogen content only needs to attain its critical value in the integration point closest to the crack tip. However, this can be different to the experiments if crack growth is governed by the presence of microstructural heterogeneities that are located dozens of microns away from the crack tip such as inclusions or segregated hardened zones. This would necessarily increase the time needed for the next crack extension event. The fracture surfaces of all the specimens were imaged after testing. As illustrated in Fig. \ref{fig:fracturesurf}(a), there was no evidence of brittle, hydrogen-assisted fracture for the specimens that passed the test, with a fully ductile fracture morphology being observed. This image was purposely chosen for the specimen with applied $K_I$ of 30 MPa$\sqrt{\text{m}}$ that passed the 7\% H$_2$S condition, where it was not possible to accurately obtain an experimental threshold $K_I$ value. On the other hand, the specimens that failed the SENT test show the hydrogen-assisted quasi-cleavage morphology typical of low-alloy martensitic steels \cite{Cho2021, Guedes2020, CupertinoMalheiros2022}, as seen in Fig. \ref{fig:fracturesurf}(b). These images are consistent with the argument above; only 10 h are needed for the first crack growth event to occur, and the sample will not fail if the peak in the steady state hydrogen concentration is not high enough, yet the failure of inner martensite interfaces is still compatible with sub-critical crack growth involving heterogeneous microstructural features. Based on this insight, a potentially suitable protocol would be to arrest the test after one day and, if the sample has not failed, conduct a subsequent three-point bending test in air to examine the fracture surface. If no cleavage-like features are observed, the applied load can be deemed to be lower than $K_{\text{th}}$; otherwise, the sample has to be considered as a failed one, $K_I \geq K_{\text{th}}$.\\ 

%-------------
%-------------

The combination of modelling and experiments also seems to suggest that first-order approximations for $K_{\text{th}}$ can be obtained for a given solution based on sub-surface hydrogen concentration (i.e., only with permeation experiments), provided that a toughness versus hydrogen content database exists. The source of the $K_{Ic}$ vs $C$ data is important when trying to account for triaxiality effects. In Refs. \cite{Cancio2010,Vera1997}, the works used to generate the database used here, a Double Cantilever Beam (DCB) geometry was used. This geometry is widely used, as it is the case of CT or SENB configurations, as they result in very high levels of crack tip constraint (positive $T$-stress \cite{Betegon1991}) and thus lower (conservative) estimates of $K_{\text{th}}$ for a given geometry. Finite element modelling can be used to quantify the role of triaxiality on crack tip fields and $K_{\text{th}}$. However, the results shown here reveal that aggressive H$_2$S solutions result in hydrogen contents that are sufficiently high to approach the plateau of the $K_{Ic}$ vs $C$ curve of low-alloy steels, where stress triaxiality plays no role.

%=============================================
\section{Conclusions}
\label{sec:conclusions}
%=============================================
We have conducted electrochemical permeation and constant-load Single-Edge Notch Tension (SENT) experiments to gain insight into the hydrogen-assisted cracking of low alloy steels exposed to corrosive environments. In addition, to gain further insight, these experiments have been combined with numerical modelling, using a phase field-based coupled deformation-diffusion-fracture model endowed with a phenomenological toughness versus hydrogen concentration degradation law. The focus was on assessing the suitability of SENT testing to characterise the susceptibility to sulfide stress cracking and other hydrogen-assisted fracture phenomena, as SENT experiments have recently attracted interest due to their ability to mimic the constraint conditions of cracked pipelines. The main findings include:\\

\begin{itemize}
    \item Permeation experiments enable quantifying hydrogen absorption as a function of the environment (H$_2$S content) and reveal the role that corrosion products have in reducing sub-surface hydrogen concentration.\\ \vspace{-10pt}

    \item The numerical model is shown to deliver reliable predictions across a wide range of applied loads, environments and geometries.\\ \vspace{-10pt}
    
    \item Laboratory and computational SENT tests enable determining a critical stress intensity factor threshold $K_{\text{th}}$ for each environment. However, this is compromised by data scatter for the low H$_2$S content scenario. The observed dispersion is attributed to the higher sensitivity of fracture toughness to hydrogen content at low hydrogen concentrations.\\ \vspace{-10pt}

    \item For severe environments, hydrogen uptake is such that $K_{\text{th}}$ is close to the toughness saturation value, the lowest value of $K_{Ic}$ in hydrogen-containing environments. As a result, triaxiality effects do not play a role and there is no advantage in using SENT testing over other tests with higher levels of crack tip constraint (DCB, CT, SENB).\\ \vspace{-10pt}

    \item The hydrogen peak near the crack tip reaches 90\% of its maximum value (the steady state result) after 10 h. This is a shorter period than the time that it takes for corrosion products to form at the surface and hinder hydrogen uptake. As such, one can neglect the role of corrosion film formation and establish a suitable testing time of less than a day. However, failure at close to 200 h has been observed for the low H$_2$S scenario and this is attributed to intermittent crack growth and its dependency on diffusion.
\end{itemize}

\section*{Declaration of Interest}
The authors declare that they have no known competing financial interests or personal relationships that could have appeared to influence the work reported in this paper.

\section*{Acknowledgements}

The authors acknowledge financial support from the EPSRC (grant EP/V009680/1). E. Mart\'{\i}nez-Pa\~neda was additionally supported by an UKRI Future Leaders Fellowship [grant MR/V024124/1].

%%%%%%%%%%%%%%%%%%%%%%%%
%\appendix
% \renewcommand*{\thesection}{\Alph{section}}

% %%%%%%% BIB files (bibtex)
%\bibliographystyle{plainnat}

% if biblatex is used
%\printbibliography

\end{document}